\documentclass[12pt]{article}

\usepackage{epsfig}

\begin{document}

\newcommand{\beq}{\begin{equation}}
\newcommand{\eeq}{\end{equation}}
\newcommand{\beqa}{\begin{eqnarray}}
\newcommand{\eeqa}{\end{eqnarray}}

\def\ov{\overline}
\def\onlyif{\rightarrow}

\def\openone{\leavevmode\hbox{\small1\kern-3.8pt\normalsize1}}

\def\a{\alpha}
\def\b{\beta}
\def\g{\gamma}
\def\r{\rho}
\def\minus{\,-\,}
\def\eks{\bf x}
\def\kay{\bf k}

\def\ket#1{|\,#1\,\rangle}
\def\bra#1{\langle\, #1\,|}
\def\braket#1#2{\langle\, #1\,|\,#2\,\rangle}
\def\proj#1#2{\ket{#1}\bra{#2}}
\def\expect#1{\langle\, #1\, \rangle}
\def\trialexpect#1{\expect#1_{\rm trial}}
\def\ensemblexpect#1{\expect#1_{\rm ensemble}}
\def\kpsi{\ket{\psi}}
\def\kphi{\ket{\phi}}
\def\bpsi{\bra{\psi}}
\def\bphi{\bra{\phi}}

\def\ditto{\rule[0.5ex]{2cm}{.4pt}\enspace}
\def\th{\thinspace}
\def\ni{\noindent}
\def\thirty{\hbox to \hsize{\hfill\rule[5pt]{2.5cm}{0.5pt}\hfill}}

\def\set#1{\{ #1\}}
\def\setbuilder#1#2{\{ #1:\; #2\}}
\def\Prob#1{{\rm Prob}(#1)}
\def\pair#1#2{\langle #1,#2\rangle}
\def\Id{\bf 1}

\def\dee#1#2{\frac{\partial #1}{\partial #2}}
\def\deetwo#1#2{\frac{\partial\,^2 #1}{\partial #2^2}}
\def\deethree#1#2{\frac{\partial\,^3 #1}{\partial #2^3}}

\newcommand{\xx}{{\scriptstyle -}\hspace{-.5pt}x}
\newcommand{\yy}{{\scriptstyle -}\hspace{-.5pt}y}
\newcommand{\zz}{{\scriptstyle -}\hspace{-.5pt}z}
\newcommand{\kk}{{\scriptstyle -}\hspace{-.5pt}k}
\newcommand{\sx}{{\scriptscriptstyle -}\hspace{-.5pt}x}
\newcommand{\sy}{{\scriptscriptstyle -}\hspace{-.5pt}y}
\newcommand{\sz}{{\scriptscriptstyle -}\hspace{-.5pt}z}
\newcommand{\sk}{{\scriptscriptstyle -}\hspace{-.5pt}k}

\def\openone{\leavevmode\hbox{\small1\kern-3.8pt\normalsize1}}

\title{Authentication and routing in simple Quantum Key Distribution networks}
\author{Andrea Pasquinucci
\\
\small
{\it {\rm UCCI.IT}, via Olmo 26, I-23888 Rovagnate (LC), Italy 
}}
\date{June 02, 2005}
\maketitle

\abstract{We consider various issues which arise as soon as one tries
to practically implement simple networks of quantum relays for QKD. 
In particular we discuss authentication
and routing which are essential ingredients of any QKD network.
This paper aims to address some gaps between quantum and networking
aspects of QKD networks usually reserved to specialist in physics
and computer science respectively.} 
\vspace{1 cm} 
\normalsize

\section{Introduction}

Until now quantum key distribution (QKD) \cite{BB84,geneva} 
has mainly been considered a point-to-point link between Alice and Bob.
However, recently, together with the arrival on the market of
the first commercial products, discussion has started on how to form 
quantum key distribution networks. 
Here we are not considering theoretical networks where one can
imagine using distributed multipartite entanglement and prolong the
distance with entanglement swapping, but practical networks which can be
implemented with todays technology.

In this paper we consider networks such as those introduced in ref.\ 
\cite{relay} (see also ref.\ \cite{oldrelay} for some unpublished
results), where a trusted quantum relay is basically
performing the well-known intercept/resend eavesdropping strategy
\cite{expcryp}, but is cooperating with Alice and Bob.
By means of such trusted quantum relays, 
it is possible to create simple networks for Quantum Key Distribution. 

In this paper we consider two, more practical, issues which arise as soon
as we form such a QKD network. The first is the problem of {\sl Authentication}.
Indeed QKD requires Alice and Bob, and in our case also the trusted relay,
to exchange messages on a classical channel. We will discuss the possibilities
and different strategies which arise in authenticating such classical
communications.

The second issue is that of {\sl Routing}. This follows from the possibility
of having networks in which QKD could be run
by using more than one intermediate relay. Different routes, i.e.\ 
intermediate relays, could exist to reach Bob, and there must exist 
a way of choosing the more appropriate among them.

This paper aims to bridge two different fields, which often adopt
different terminologies, that of quantum physics, and in particular
quantum information theory and quantum cryptography, 
and that of computer science, in particular networking and its security.
As such an effort has been done to adopt a language understandable by both
and to blend results from the two fields.

The paper is organized as follows: In section 2, we recall the 
protocol introduced in ref.\ \cite{relay} and discuss various issues 
related to the authentication of the classical channel.
In section 3, we discuss some very basic issues of routing in QKD
networks indicating the main problems which we believe are still open.
In section 4 we point out how the initial authentication of the participants
to a QKD network, including the relays, could be implemented in different
ways depending on the security requirements imposed.
In section 5 we conclude with some final remarks.

\section{QKD networks and authenticity of the classical channel}

Let's start by considering the simplest QKD setup with a trusted quantum relay.
In the simplest protocol (see ref.\ \cite{relay}) the relay acts practically 
doing an intercept/resend strategy to forward qubits\footnote{A computer
scientist can safely assume that a qubit is practically realized with
a single photon.} 
sent by Alice to Bob. 
At the end Trent, the relay, announces to Alice and Bob the basis he 
has used for each qubit. His participation in the protocol ends here. 

Obviously, Trent, Alice and Bob need to use a classical channel of 
communication which guarantees to them
that messages exchanged between all three are not modified by Eve nor
that Eve can inject fake messages in the channel.\footnote{Usually it
is assumed that Eve can attack in any way she wants the QKD quantum channel but
that she can only passively eavesdrop on the QKD classical channel.} 
We say that the 
messages sent on the classical channel by Alice, Bob and Trent are 
authenticated and integral. {\sl Authenticity} and {\sl Integrity} 
of a classical communication channel can be obtained with standard 
classical cryptographic algorithms. For our purposes it is important to 
stress that Authenticity requires Alice, Bob and Trent to share 
some secret keys to prove that they are the source of the information.
For this reason often QKD is called a {\sl key extension protocol}, since
to run QKD Alice and Bob, and in our case also Trent, need to start with
some secret keys already shared. Usually the new key produced by QKD
will be partly used to encrypt the secret data that Alice (Bob)
wants to send to Bob (Alice),
and partly used to refresh the shared keys for the authentication.

We want to stress a point which can lead to some confusion. In our
setup, a Quantum Key Distribution protocol
is run to create in a secure way a {\sl secret key} shared by 
Alice and Bob. This secret key will be then used by Alice (Bob) 
to encrypt some data for Bob (Alice) using classical algorithms like 3DES, 
AES or OTP. Thus, from our point of view, QKD is used only to create such
a secret key and we will not discuss how such a key will be used by
Alice and Bob. In running a QKD protocol, Alice, Bob and Trent need to 
exchange some information on a classical channel, and in this paper when we 
discuss classical communications between them, we always consider the
communications needed to run the QKD protocols, and not the exchange
of secret data between Alice and Bob.

To run the protocol, Alice, Bob and all Trents need to share beforehand
some secret authentication keys. These keys need to be substituted after
their use, how often depends on the classical authentication algorithm 
adopted. To refresh the authentication keys, Alice, Bob and all Trents 
can use part of the keys created by running QKD. As we will see, in our
case the keys generated by QKD between two participants, one of which is a 
Trent, can be used fully to refresh the corresponding authentication keys.
Instead the authentication key between Alice and Bob should be extracted
(subtracted) from their final secret key.

An important point in practical implementations is that the rate of 
refreshment of the authentication keys must be compatible with the rate
of creation of new keys by QKD. In our case this means that for
authentication keys involving at least one Trent, the rate of refreshment
must be at most equal to the rate of creation of the corresponding 
QKD key. Instead the rate of refreshment of the authentication 
key shared between Alice and Bob, must be much lower than the rate of
creation of the corresponding QKD key, otherwise there will not be any
key left that Alice and Bob can use to encrypt their secret data.
In this paper we will not consider any particular implementation
of the authentication scheme.

Thus in our situation, there are two possible cases for the secret 
authentication keys shared between the participant before QKD is started:
\begin{enumerate}
\item Alice and Bob share each a secret authentication key with Trent
\item Alice and Bob share each a secret authentication key with Trent, 
and Alice shares a secret authentication key with Bob.
\end{enumerate}
In both cases Alice and Bob must share a secret authentication 
key each with Trent, 
and thus need to refresh these keys once they have been used. The simplest
way they can do this is by running a QKD protocol also with Trent and use
the resulting secret key to authenticate the classical channel. In ref.\ 
\cite{relay} it has been shown that by involving Trent in the error correction
and privacy amplification phases of the protocol, it is possible to create
out of the same raw data, three keys, one shared between Alice and Trent,
one shared between Trent and Bob, and one shared between Alice, Bob and 
possibly Trent.

Since Trent is now an active part in the protocol, and shares both
classical and quantum keys with Alice and Bob, he does not play anymore
the role of trusted arbitrator, but becomes a full member of the quantum
protocol. For this reason, and following ref.\ \cite{relay}, we prefer 
to call her Carol.

For clarity, we repeat here the full protocol.
\begin{enumerate}
\item Alice prepares a qubit in one of the four states $\pm x$ or $\pm
y$, and sends it to Carol
\item Carol measures, as if she was Bob, in either the $X$ or the $Y$
basis. According to the result of this measurement, she prepares the same
state that she found in this measurement and sends it to Bob.
\item Bob measures again in the $X$ or the $Y$ basis.
\item Alice, Carol and Bob repeat the first 3 steps many times.
\item Alice, Carol and Bob announce to each other
which basis they used, and they divide the qubits in four groups: 
 \begin{enumerate}
 \item the first group is the one of the qubits for which Alice, Carol and Bob 
used the same basis; Alice and Bob, on the qubits of this group, proceed with
the estimate of the error rate followed by error correction and privacy
amplification to obtain a secret key --- as in the standard BB84 protocol.
 \item  the second group is the one of the qubits
for which Alice and Carol used the same basis but Bob the opposite one;
Alice and Carol, on the qubits of this group,
proceed with the estimate of the error rate
followed by error correction and privacy
amplification to obtain a secret key.
 \item the third group is the one of the qubits for which Bob and Carol used 
the same basis but Alice the opposite one;
Bob and Carol, on the qubits of this group, proceed with
the estimate of the error rate followed by error correction and privacy
amplification to obtain a secret key.
 \item the fourth  group is the one of 
the qubits for which Alice and Bob used the same basis but Carol 
the opposite one; this group cannot be used.
 \end{enumerate}
\end{enumerate}
In this way, 3 secret keys are created, one between Alice and Bob which
Carol can reconstruct since she has the same raw data; one between Alice
and Carol, and one between Bob and Carol.

Notice that in the protocol we have presented, Carol does not participate
in the error correction and privacy amplification phases with Alice 
and Bob on the qubits of group (a). As discussed in ref.\ \cite{relay}, 
if Carol follows the public discussion between Alice and Bob, she will
be able to obtain a final secret key very similar, but usually not identical,
to the one of Alice and Bob. The reason for this are few experimental errors
or errors induced by Eve, that Carol is not able to correct. In any case,
the knowledge on the final key shared between Alice and Bob that Carol
can obtain by listening to the public discussion between them, is very
large and cannot be reduced. So it should be assumed that Carol can 
always in practice learn the secret key shared by Alice and Bob.

The secret keys Alice-Carol and Bob-Carol can be fully used to authenticate 
the communication channel. As for the keys between Alice and Bob, as we said
there are two possibilities:
\begin{enumerate}
\item Alice and Bob do not share any secret authentication key 
and rely on Carol for the
authentication; thus in this case every classical message sent by Alice to Bob
(and vice-versa), is received by Carol who verifies its authenticity and
integrity, and then sends it to Bob (or Alice) adding her own 
authentication code; notice that in this case Bob (or Alice) verifies the
authentication code of Carol, and not the one of the true sender of the 
message (see fig.\ \ref{fig:fig1}).\footnote{In this and in the following 
Figures
the ''message'' is a classical message of a phase of the QKD protocol,
for example to realize the sifting procedure, and not the secret data that 
Alice wants to send to Bob.}
\item Alice and Bob share some classical secret authentication key 
and use it to generate the authentication code added to
the classical messages which they exchange without any intervention by Carol;
as we said, in this case part of the final secret key created by QKD 
between Alice and Bob must be used for this task (see fig.\ \ref{fig:fig2}).
\end{enumerate}

\begin{figure}[t]
\begin{center}
\leavevmode
\hbox{%
\epsfxsize=4.5in
\epsffile{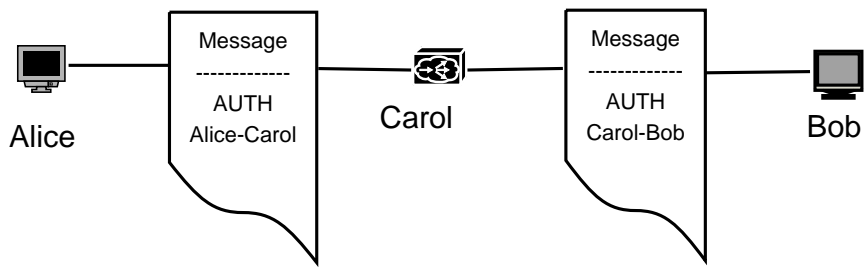}}
\caption{Authentication codes on a classical message exchanged between 
Alice and Bob when they share an authentication key
only with Carol}
\label{fig:fig1}
\end{center}
\end{figure}

\begin{figure}[t]
\begin{center}
\leavevmode
\hbox{%
\epsfxsize=4.5in
\epsffile{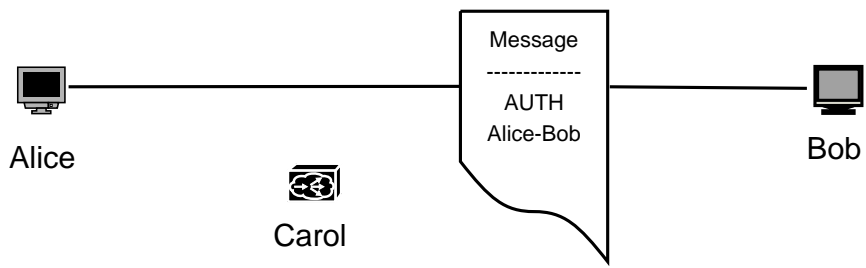}}
\caption{Authentication codes on a classical message exchanged 
between Alice and Bob when they share an authentication key}
\label{fig:fig2}
\end{center}
\end{figure}

\noindent
Of course these two cases have both points in their favor and against.
If Alice and Bob do not share any authentication key:
\begin{itemize}
\item Alice and Bob depend on Carol also for their classical communications
of the QKD protocol (notice anyway that in our QKD protocol Alice and
Bob must in any case exchange classical messages with Carol)
\item Carol has to do more work since she has to verify the authentication
code for each message she receives and add her own to the message before
sending it
\item Alice has not direct confirmation of Bob's identity, and vice-versa
\item no part of the QKD final secret key created between Alice and Bob is used 
for authenticating the classical channel
\item the level of trust on Carol by Alice and Bob does not really change,
since theoretically in any case Carol could practically learn 
the final QKD secret key shared between them.
\end{itemize}
Instead, if Alice and Bob share a secret authentication key:
\begin{itemize}
\item Alice and Bob can send directly to each other the authenticated
classical messages of the QKD protocol without need of Carol relaying 
and authenticating the messages
\item Carol has less work to do since she does not need to relay
and authenticate Alice-Bob messages
\item Alice has direct confirmation of Bob's identity, and vice-versa
\item part of the QKD key created between Alice and Bob is used
for authenticating the classical channel thus decreasing the final rate
of creation of the secret key shared by Alice and Bob
\item the level of trust on Carol by Alice and Bob does not increase,
since theoretically in any case Carol could  practically learn the 
final QKD secret key shared between them.
\end{itemize}

\section{QKD networks and routing}
As soon as there are networks, the problem of routing arises. In
the simplest case we can pose the problem as follows. Suppose that
there is a simple network with one Alice, a few Bobs and one Carol. 
Alice and the Bobs are all connected to Carol. How can Alice inform
Carol to which Bob she wants to connect?

With more complex networks, not only Alice should tell the Carol she 
is connected to, but the first Carol should also know to which second
(third etc.) Carol the final Bob is connected to. 
Thus it is necessary to have at least three elements:
\begin{enumerate}
\item a classical communication protocol between Alice, Carol and Bob 
to be able to exchange the informations needed to establish 
the {\sl route\/} between them
\item some kind of {\sl routing tables\/} held by Carol which permits
her to know how to reach all Bobs
\item a dynamical protocol which allows
to modify these tables when Bobs are added/removed to/from the network.
\end{enumerate}

\noindent
It could be useful to add a comparison with a network well known to everybody,
the public Internet. When for example we {\sl surf on the web} the following
happens:
\begin{enumerate}
\item the {\sl name} of the web site we want to reach, for example
{\tt lanl.arxiv.org}, is translated
in the address {\tt 204.121.6.57} 
of the web server using a protocol called {\sl Domain
Name System} (DNS) \cite{DNS}
\item the request for the page is sent to the address of the web server
using the {\sl Internet Protocol} (IP) \cite{IP}, this protocol specifies how
addresses are formed and how data is formatted and transmitted
\item the IP protocol also specifies the form of the routing tables and how
data is routed by the {\sl gateways} or {\sl routers}, i.e.\ the devices 
which play the role of the relay between the sender and the receiver
of the data
\item the {\sl Border Gateway Protocol} (BGP) \cite{BGP} specifies how the
routing tables are updated in real-time so that all gateways/routers are 
always able to send the data along the shortest and loop-free path.
\end{enumerate}

\noindent
For QKD networks like the ones described in this paper and in ref.\ 
\cite{relay}, it will be necessary to adopt or create protocols to 
fulfill similar tasks. The IP and related protocols cannot be adopted 
as they are, since they lack practically any security feature. It is
instead fundamental that all communications and informations exchanged
in a QKD network are trusted and verified.

A detailed formulation of protocols of this kind is beyond the scope
of this paper. Below instead we will make some general considerations 
on the properties and characteristics of such protocols.

First of all, the main goal of such protocols is to allow 
Alice and Bob to create
a secret key using a QKD protocol. Thus Alice and Bob should be
certain of the identity of each other. Moreover, since the relays 
we use can practically learn the secret keys, they must trust them.
So in a large network, Alice and Bob could require to accept or select
the relays through which their communications, both quantum and classical,
will pass. 

Here we assume for simplicity that both quantum and classical
communications between Alice and Bob follow the same route and pass 
through the same relays. This assumption seems to be reasonable 
when considering networks since it is certainly impractical to have
different networks for the quantum and classical phases of the QKD protocol.
Anyway there could be situations where the classical and quantum data
exchanged between Alice and Bob follow two different paths; we will not 
consider further this scenario in this paper.

We should stress that Alice and Bob can not verify a priori
through which relays their qubits will pass. Indeed a relay could route
the classical communications through one path and the qubits through a 
different one without Alice and Bob noticing it at the moment. Of course,
since the relays participate in the sifting procedure, Alice and Bob
will realize a posteriori which route the qubits have followed by checking 
which relays are announcing the basis chosen for each qubit.

\begin{figure}[t]
\begin{center}
\leavevmode
\hbox{%
\epsfxsize=4.5in
\epsffile{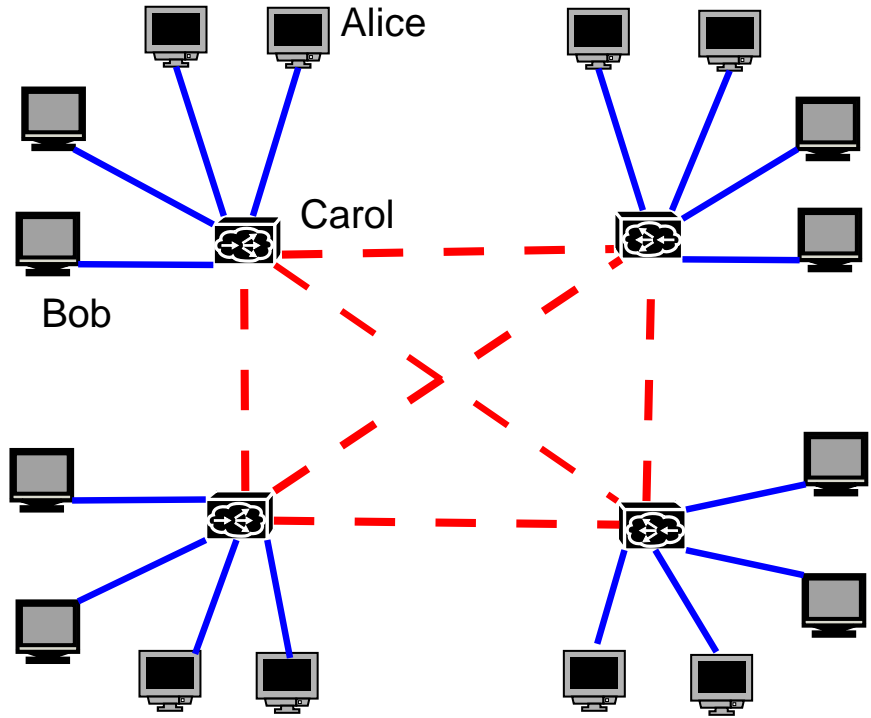}}
\caption{A simple network with Star topology between Alices and Bobs with
Carols as centers, and Full-Mesh between Carols}
\label{fig:fig3}
\end{center}
\end{figure}

To make our discussion more practical, 
let's consider the network topology described in ref.\ \cite{relay} and
in Figure \ref{fig:fig3}.
In this case all Carols are connected to each other, thus Alice to reach
any Bob should pass through one or two Carols. We consider
the case in which the path between Alice and Bob is Alice, Carol-1, Carol-2 
and Bob.

As we have discussed in the previous section, Alice always shares 
an authentication 
key with Carol-1 to authenticate the classical channel between them, and
Bob shares an authentication key with Carol-2. 

Analogously Carol-1 and Carol-2 share an authentication key that they use to 
authenticate their communications. In particular the two Carols need to
inform each other of the Alices and Bobs that are connected to them.
Indeed the {\sl routing tables} held by our Carols contain the list of
Alices and Bobs connected to each one of them, and all Carols need to 
exchange classical communications to keep these tables updated.

As in the previous section, the refreshing of the authentication keys shared
between any two members of the QKD network, can be extracted from parts
of the raw data not used to form the QKD secret key between Alice and Bob.
In our case study, 
from the raw data it can be extracted: an authentication key 
between Alice and Carol-1, an authentication key between Alice and Carol-2, 
an authentication key between Carol-1 and Carol-2,
an authentication key between Carol-1 and Bob, an authentication key 
between Carol-2 and Bob, plus the key between Alice and Bob. 
Moreover, if needed, the authentication key between Alice and Bob must 
be extracted from their QKD final key. 
In total 7/8 of the raw data is used to create secret keys.

Consider then when Alice wants to start a QKD with Bob. First Alice has 
to establish a route to Bob, and Bob has to accept to run the protocol. 
Thus Alice sends her request, an authenticated classical message, to Carol-1.
Carol-1 first checks in her routing tables if Bob is directly connected to her,
otherwise she finds to which Carol is Bob connected and sends the message 
to Carol-2 who forwards it to Bob.
Bob should reply, accepting to run a QKD protocol with Alice, sending
an authenticated classical message back through the same route.
In the same way the participants exchange authenticated classical messages 
for the synchronization of the various phases of the QKD protocol, 
the messages of sifting, error correction and privacy amplification, 
and the final messages for ending the QKD run.

\begin{figure}[t]
\begin{center}
\leavevmode
\hbox{%
\epsfxsize=5.0in
\epsffile{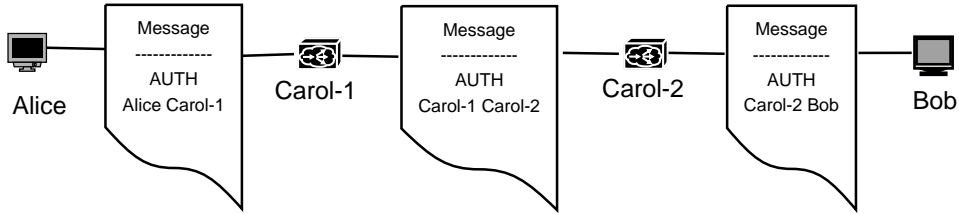}}
\caption{Step by step authentication codes on a classical message 
exchanged between Alice and Bob.}
\label{fig:fig4}
\end{center}
\end{figure}

One important point is how Alice and Bob should authenticate. In the 
simplest case, Alice shares a classical key only with Carol-1, and verifies
only the authentication codes by Carol-1. This means that all classical
messages sent by Carol-2 for the sifting, and Bob through Carol-2 
to Alice, are received
by Carol-1 who verifies the authentication code and if correct, resends
them to Alice with her own authentication code. In this way, authentication
is added and verified step by step on the route, i.e.\ only locally, 
and Alice has to fully trust Carol-1 on the identity and authenticity of
Carol-2 and Bob (see fig.\ \ref{fig:fig4}). 
Thus Alice in this case needs to share a classical key
only with Carol-1.

\begin{figure}[t]
\begin{center}
\leavevmode
\hbox{%
\epsfxsize=5.0in
\epsffile{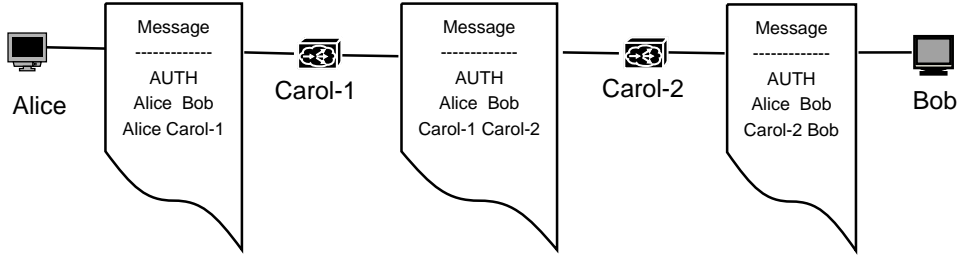}}
\caption{Authentication codes on a classical message
exchanged between Alice and Bob when they share an authentication key.}
\label{fig:fig5}
\end{center}
\end{figure}

Since at the end the important point is that Alice and Bob identify 
and authenticate each other, the previous case can be improved by
having Alice and Bob share an authentication key and verify directly
each other authentication codes (see fig.\ \ref{fig:fig5}). 

\begin{figure}[t]
\begin{center}
\leavevmode
\hbox{%
\epsfxsize=5.0in
\epsffile{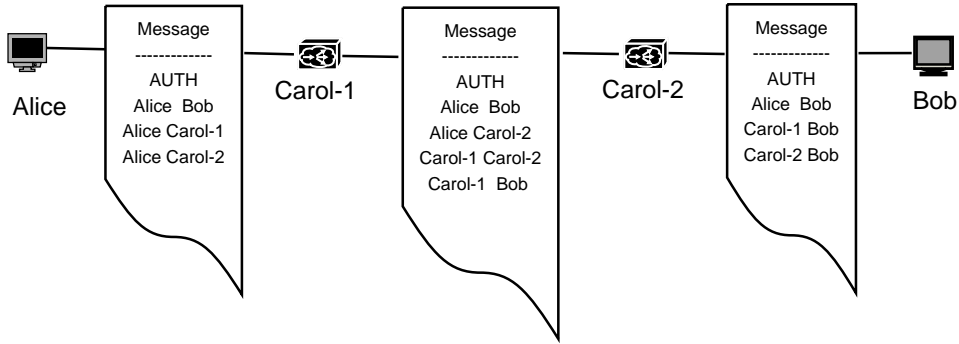}}
\caption{Authentication codes on a classical message
exchanged between Alice and Bob when everybody shares an authentication key
with everybody else.}
\label{fig:fig6}
\end{center}
\end{figure}

The most general case is the one in which Alice and Bob share a classical
key also with all Carols and are then able to verify the authentication
codes of everybody else. In practice in this case, whoever receives a
classical message, verifies all authentication codes added to it, and adds
his/her own authentication code before forwarding it (see fig.\ \ref{fig:fig6}).  

These three general situations are of increasing complexity and require
more resources in turn. 
The more complex they are, the more layers of security they add 
to the authentication
of the classical messages exchanged for routing and the phases of the
QKD protocol. 

At this level of analysis it is not clear if the simplest case is
sufficiently secure. This depends also on the level of trust and complexity
of the network since there could be more general cases in which Alice and Bob 
do not want a particular Carol to be part of the route.

\section{Initial Authentication in large networks}

We have seen in the previous sections, that our networks require Alice 
at least to share an authentication key with the Carol she is connected
to, and similarly for Bob. In principle it is not necessary for Alice
to share a key with Bob before running QKD. Indeed consider the case
of a large network, in this case all Carols have the list of all
participants, which can be easily extracted by the routing tables. In
other words, all Carols can act as directory services, some 
kind of {\sl Telephone Directory\/} in which Alice can find a Bob 
to connect to.
Using her authenticated classical channel with Carol, Alice can then ask 
her the list of Bobs available at any time. 

To start a QKD with a Bob present in Carol's directory, Alice can adopt
one of three general strategies, she can choose the one she prefers
depending on the kind of communication she needs and the level of 
security of initial identification and authentication of the corresponding
Bob. 

The first and simplest case is when Alice relies entirely on Carol, thus
she just shares a secret key with Carol and trust her completely on 
identifying and authenticating Bob (see fig.\ \ref{fig:fig4}). 
Obviously every time Alice will 
run a QKD with the same Bob, she will trust entirely Carol and have no 
direct proof that she is communicating with the same Bob as the previous
time.

The simplest improvement on this approach, which is adopted by many protocols
in classical cryptography, is to trust Carol for the connection to Bob 
only the first time. Thus with the first run of the QKD protocol, Alice
and Bob create a shared secret key but they do not use this key to 
encrypt some data, but they keep it as their authentication key. When
Alice and Bob run again a QKD protocol they share an authentication key
and they can authenticate directly each other. Thus in this case the 
first run of QKD has the special purpose
to create a shared key for future authentications. In this case, the 
first run of the QKD is like the one in fig.\ \ref{fig:fig4}, whereas all the others
like the one in fig.\ \ref{fig:fig5}.

Obviously, this can be done not only between Alice and Bob, but also between
Alice and Carol-2, and Bob and Carol-1.\footnote{We assume that Carol-1 and 
Carol-2 already share an authentication key used to authenticated 
the messages they exchange for the routing.}
Thus the first run of the QKD protocol
can be used entirely to create authentication keys between all participants
in the communications so that after the first run of QKD 
the authentications codes can be like the ones in fig.\ \ref{fig:fig6}.

The last approach to the initial distribution of authentication keys,
is the one where Alice shares an authentication key with Bob, and in case
with all other Carols,
before starting any QKD. This key must be exchanged on a different channel,
like person to person exchange of the key on a paper slip. Of course
this case requires much more work for Alice to enlarge her QKD network, 
since the initial authentication phase requires some kind of out-of-band 
communication.

Notice that combining the possibility of creating shared authentication
keys, with the possibilities of authentication of the classical messages
exchanged during the QKD protocol described in the previous section, 
Alice and Bob have the possibility of fine tuning the level of control
they want to apply on their classical communications. Notwithstanding
all this, Alice and Bob must anyway trust all Carols, since in any case 
all Carols acting as relay for their communications 
could always practically learn the secret key that Alice and
Bob generate with QKD.

\section{Some final remarks}

One of the peculiarity of QKD, is that Alice and Bob have different 
hardware boxes. In the simplest case, Alice sends a qubit (photon) whereas
Bob receives it. Thus in some current implementations, Alice has a laser 
whereas Bob has a detector.\footnote{In the Plug\&Play scheme both the laser
and detector are at Bob's site; in any case still Alice and Bob boxes
are different.}
In a network like the one in Figure 3, all Carols must implement both 
kinds of hardware, a Bob-like box facing Alice, and an Alice-like box
facing Carol-2 or Bob. 

If Alice has only an Alice-like box, it is impossible for her with 
the protocols considered in this paper, to run a 
QKD protocol with any other Alice in the network. The easy solution to
this is to provide all Alices and Bobs with a double box, 
containing both the hardware
of an Alice-like box and Bob-like box and with the possibility of acting
as either of the two depending on the setup of the connection.
In this way any participant in the network can run a QKD protocol with
anyone else.

Since in any implementation of the BB84 protocol, the role of Alice and
of Bob is similar, the presence of the trusted relays allows also
Alice and Bob to have boxes from different manufacturers,
adopting different implementations or distribution platforms (optical
fibers or free space).

Finally, Alice can also create keys with two or more Bobs at the same time. 
The simplest way is to alternate (i.e.\ multiplexing), 
either qubit by qubit or run by run of the QKD protocol,
the destination of the qubit. This can be done by time or wavelength 
multiplexing, that is by integrating the QKD networks discussed in this paper
with multi-user QKD schemes which have been proposed to allow
a single Alice to be directly connected to more than one Bob without
any relay \cite{multiu}.
Of course in our case the multiplexing of the qubits sent by Alice 
for the various Bobs is coupled to the routing done by either 
Carol-1 or Carol-2, who should deliver the correct qubit to each Bob.


\section*{Acknowledgments}
We thank H.\ Bechmann-Pasquinucci for inspiring remarks.

\end{document}